\documentclass[dvips,aoas,submission]{imsart}

\RequirePackage[OT1]{fontenc}
\RequirePackage{amsthm,amsmath}
\usepackage{amsfonts}
\usepackage[dvips]{graphicx}
\RequirePackage[]{natbib}
\RequirePackage[colorlinks,citecolor=blue,urlcolor=blue]{hyperref}

\usepackage{algorithm}
\usepackage{algorithmic}
\newcommand{\myiid}{\stackrel{\mathrm{iid}}{\sim}}

\usepackage{array}
\usepackage{changepage}
\usepackage{color, soul}
\usepackage{multirow}
\usepackage{enumerate}
\usepackage{etoolbox}
\arxiv{stat.AP/0000000}

\startlocaldefs
\numberwithin{equation}{section}
\theoremstyle{plain}

\endlocaldefs

\begin{document}
\begin{frontmatter}
\title{A semi-automatic method to guide the choice of ridge parameter in ridge regression} 
\runtitle{Choice of ridge parameter in ridge regression}

\begin{aug}
\author{\fnms{Erika} \snm{Cule}\thanksref{m1}\ead[label=e1]{erika.cule05@imperial.ac.uk}}
\and
\author{\fnms{Maria} \snm{De iorio}\thanksref{m2}\ead[label=e2]{m.deiorio@ucl.ac.uk}}

\runauthor{E. Cule and M. De Iorio}

\affiliation{Imperial College London\thanksmark{m1} and University College London \thanksmark{m2}}

\address{Department of Epidemiology and Biostatistics \\
School of Public Health \\
Faculty of Medicine \\
Imperial College London \\
St Marys Campus \\
Norfolk Place \\ 
London W2 1PG \\
\printead{e1}}

 \address{Department of Statistical Science\\
 University College London\\
 Gower Street\\
 London WC1E 6BT\\
 \printead{e2}}
\end{aug}

\begin{abstract}
We consider the application of a popular penalised regression method, Ridge Regression, to data with very high dimensions and many more covariates than observations. Our motivation is the problem of out-of-sample prediction and the setting is high-density genotype data from a genome-wide association or resequencing study. Ridge regression has previously been shown to offer improved performance for prediction when compared with other penalised regression methods. One problem with ridge regression is the choice of an appropriate parameter for controlling the amount of shrinkage of the coefficient estimates. Here we propose a method for choosing the ridge parameter based on controlling the variance of the predicted observations in the model. 

Using simulated data, we demonstrate that our method outperforms subset selection based on univariate tests of association and another penalised regression method, HyperLasso regression, in terms of improved prediction error. We extend our approach to regression problems when the outcomes are binary (representing cases and controls, as is typically the setting for genome-wide association studies) and demonstrate the method on a real data example consisting of case-control and genotype data on Bipolar Disorder, taken from the Wellcome Trust Case Control Consortium and the Genetic Association Information Network. \end{abstract}



\end{frontmatter}

\section{Introduction}

We are interested in the problem of out-of-sample prediction in a high-dimensional regression setting, when data consist of many more predictor variables than observations. Our motivation is the prediction of a phenotype (observable characteristic) of interest using an individual's genetic information, and possibly other covariates. In this introduction we describe the problem of phenotype prediction using genetic data, and we outline the specific issue, selection of the ridge parameter in Ridge Regression (RR, \cite{Hoerl:1970p25}), that we address in this paper.

\subsection{Genetic Risk Prediction}

The data that motivate this paper arise in contemporary research into the relationship between genetic and phenotypical variation. Modern Genome-Wide Association Studies (GWAS) investigate a phenotype of interest, which may be a continuous trait (such as blood pressure) or (more commonly in GWAS) a binary trait describing a disease case (1) or disease-free control (0). The corresponding genetic data consist of typed genetic variants (normally single nucleotide polymorphisms (SNPs)) for each individual. Current technology allows the typing of thousands to hundreds of thousands of variants directly, and this number may be augmented to millions through imputation. These genetic variants are densely spaced along the genome and are highly correlated due to the phenomenon of linkage disequilibrium (LD), the co-inheritance of nearby genetic variants that results in their frequencies deviating from those that would be expected if each SNP was inherited independently. 

The phenotypes can be viewed as the outcome variable in a regression setting, with the genetic variants as the predictors. As genetic variants are fixed at birth, but phenotypes may not present until later in life, there is the potential for genetic data to be used to predict future phenotype. In a clinical setting, such a prediction could suggest lifestyle modifications or pharmaceutical interventions that would prevent or delay the onset of disease. Then, our aim is to construct a model to predict the phenotype of interest using the genotype data from individuals for whom the (potentially future) disease state is unknown. 

RR [\cite{Hoerl:1970p25}] is a means of estimating regression coefficients when data are high-dimensional and/or contain correlated variables. This is often the case in genetic data, as described above. RR can be used to obtain stable parameter estimates when standard multiple regression approaches would fail. Among a number of regularized regression methods, RR was shown to perform best in terms of predictive performance [\cite{Frank:1993p260}]. By making use of an orthogonal transformation of the high-dimensional data, RR coefficients can be computed in a manner that is computationally efficient. In this paper we demonstrate that RR has good predictive performance in high-dimensional settings. In addition, we propose a method for choosing the penalty parameter when there are more predictors than observations. 

Our focus here is on prediction. Much of the analysis of genetic data to date has focused on the identification of causal variants. Using GWAS data with case-control phenotypes, statistical differences in genotype at a certain genetic locus in populations of cases and matched controls is seen as indicative of association. A typical approach is a ``single-SNP'' analysis, in which each SNP is tested individually for association with the outcome variable. Due to the large number of tests, a stringent threshold for significance is required. SNPs that are found to be associated with outcome at some significance threshold are viewed as either affecting the outcome, or likely to be correlated with such a SNP. In the latter case, further SNPs near the associated SNP are typed in an attempt to identify the causal variant(s). Once identified, these genetic variants can give insight into the biological mechanisms of the disease and may indicate potential drug targets. 

Existing attempts to use genotype information to predict disease risk have taken causal variants identified in previous studies and used these to construct a predictive model. One approach is to base the increase in disease risk on a cumulative count of known risk variants [\cite{Meigs:2008p13}]. The model may take into account the estimated effect size of each variant [\cite{Karlson:2011p2510}]. To date, the improvement in prediction of disease risk based on genetic data over traditional risk prediction models (using clinical data and family history) has been too small to have clinical utility [\cite{Talmud:2010p80}]. Whilst genome-wide association studies have collectively identified thousands of genetic variants associated with diverse phenotypes [\cite{Hindorff:2009p938}], overall, effect sizes of identified variants are small and much heritability remains unexplained [\cite{Eichler:2010p2504}]. Suggested sources of this so-called ``missing heritability'' include variants whose effect size does not meet the stringent significance threshold required in univariate tests, rare variants of large effect which current studies are underpowered to detect, or the combined effect of multiple SNPs whose individual effects do not reach univariate significance. 

The single-SNP tests described above do not take into account the combined effect of multiple SNPs. However, ordinary multiple regression cannot be applied to genetic data due to the high dimensionality of the data and the correlation among SNPs. Previous work on the analysis of all SNPs simultaneously [\cite{Ayers:2010p2075,Hoggart:2008p147}] focused on the identification of causal variables when genotypes are in LD.  The penalised regression methods used in these studies estimate some regression coefficients (effect sizes of variables) to be exactly zero and remove the corresponding variables from the model. This results in the information in the removed variables being lost if the estimated effect size falls below the effective significance threshold determined by the penalty parameters in the penalised regression. RR, in contrast, estimates the effect size of all the variables. We suggest that this will improve predictive performance [\cite{Kooperberg:2010p1145}]. The estimated effect sizes of non-causal variants will be centred around zero and will cancel each other out, whereas the true but small effects will be retained in the model. 

\subsection{The choice of shrinkage parameter in Ridge Regression}\label{sec:intro:choice}
RR was proposed as a means of estimating regression coefficients with smaller mean-square error than their least squares counterparts when predictors are correlated [\cite{Hoerl:1970p25}]. RR has been extended to the logistic regression model [\cite{Lee:1988p928,Cessie1992}] and has been applied to genetic data when predictor variables are in high LD with the aim of improving identification of causal SNPs [\cite{Malo2008}]. \cite{Cule:2011p2590} proposed a test of significance of regression coefficients estimated using RR, facilitating the use of RR to guide variable selection in the analysis of genetic data. In common with other shrinkage methods such as the LASSO [\cite{Tibshirani1996}] or the elastic net [\cite{Zou:2005p597}], RR requires the specification of a penalty parameter that controls the degree of shrinkage of the coefficient estimates. A number of methods have been proposed to estimate this parameter in RR based on the data, but no consensus method provides a universally optimum choice. \cite{Hoerl:1970p25} proposed a graphical method, the ridge trace. Based on the test of significance of ridge regression coefficients, \cite{Cule:2011p2590} introduced the $p$-value trace, a plot of the significance of RR coefficients with increasing shrinkage parameter. The $p$-value trace can be used to evaluate the changing significance of the regression coefficients with increased shrinkage parameter. However, graphical methods rapidly become unfeasible when the number of predictors is large. Data-driven methods have also been proposed. Some popular choices are those of \cite{Hoerl1975} and \cite{Lawless:1976p100}. However, both of these rely on least-squares estimates of the parameters, which do not exist when the number of variables in the model exceeds the number of observations. 

Cross-validation is another approach that has been proposed to choose the ridge parameter [\cite{Golub:1979p86}]. However, this has the disadvantage of being computationally intensive. Techniques corresponding to those proposed for linear RR have been used in logistic RR [\cite{Vago:2006p151}]. A simulation study investigated the performance of various logistic ridge estimators; however, the simulations  only consider regression problems with fewer predictors than observations, and do not conclude that a single ridge estimator is best in terms of reducing mean squared error in all regression situations. 

In this paper, we propose a semi-automatic method to guide the choice of ridge parameter in RR. Our proposed method is valid when the regression problem comprises more predictors than observations. The method proposed here is motivated by two observations:

The first observation is that the ridge estimator proposed by \cite{Hoerl1975} is based on the ``ideal'' ridge estimator. This ``ideal'' estimator is defined through the true (but unknown) regression coefficients, and is derived by first transforming the model into canonical form. In their ridge estimator, \cite{Hoerl1975} replace the unknown regression coefficients in the ``ideal'' estimator with their least squares counterparts. In data, such as genetic data, with more predictors than observations, these least squares estimates do not exist. However, due to the correlation structure in genetic data, it is likely that most of the variance in the predictors is explained by the first few components when the model is transformed to canonical form. By transforming the model in this way and using only the first few principal components (PCs) in the estimator of \cite{Hoerl1975}, we can feasibly construct a ridge parameter based on the meaningful component directions in a genetic data set.

The second motivation is arrived at by considering the relationship between RR and principal components regression (PCR, \cite{Hastie:2009p71}). RR coefficients can be derived by shrinking the coefficients from a PCR including all the components, in proportion to their corresponding eigenvalues. Coefficients corresponding to components which explain a smaller proportion of variance in the data are shrunk more. \cite{Hastie1990} define the degrees of freedom for variance of a penalised regression fit. In PCR, the degrees of freedom for variance are equal to the number of components retained in the model; in RR, the degrees of freedom for variance depend on the ridge parameter. By restricting the degrees of freedom for variance in a ridge regression fit to be the same as for a corresponding PCR we can develop a model with the same prediction variance as a PCR. In situations that commonly arise in GWAS, this results in a model with smaller bias than the corresponding PCR. This is because whilst PCR discards the information in the components not used in the model, RR `spreads out' the shrinkage across all components, thus making more efficient use of the degrees of freedom for variance. The components that explain less of the variance in the data are penalised more. The result is a model with smaller average prediction error. 

The remainder of this paper is organised as follows: In Section \ref{sec:ridgeestimator} we introduce RR and derive our proposed ridge estimator. In Section \ref{sec:simstudy} we illustrate aspects of our approach using simulation studies. In Section \ref{sec:realdata} we present an evaluation of our method using data on Bipolar Disorder from the Wellcome Trust Case-Control Consortium (WTCCC) [\cite{WTCCC2007}] to construct a prediction model. This model is then evaluated using data from the Genetic Association Information Network (GAIN) [\cite{GAIN}]. In Section \ref{sec:compissues} we discuss some computational issues that arise when fitting a ridge regression model to very high-dimensional data. 
\section{Ridge estimator}
\label{sec:ridgeestimator}
\subsection{Regression}
Consider a regression problem in which each data point $(Y_i, \mathbf{x}_i), \hspace{1pt} i = 1, \dots, n$,  comprises a response variable $Y_i$ and a corresponding vector of $p$ predictors, $\mathbf{x}_i = \left( x_1, \dots, x_p\right)$. In the linear regression model, the relationship between $n$ such responses and predictors is given by:
\begin{equation}\label{eq:linmodmat}
\mathbf{Y} = \mathbf{X}\boldsymbol{\beta} + \boldsymbol{\epsilon}
\end{equation}
Here, $\mathbf{Y} = \left( Y_1, \dots, Y_n\right)$ and $\mathbf{X}$ is an $n \times p$ matrix with rows $\mathbf{x}_i$. $\boldsymbol{\beta} = \left( \beta_1, \dots, \beta_p \right) $ is a vector of regression coefficients, and $\boldsymbol{\epsilon} = \left( \epsilon_1, \dots, \epsilon_n \right)$ is a vector of independent and identically distributed normally distributed random errors, $\mathbb{E}\left( \epsilon_i \right) = 0$ and $\mathbb{E}\left( \epsilon_i^2 \right) = \sigma^2$. When $\mathbf{X}$ is of full rank, the least-squares estimators of the $p$ parameters $\beta_1 \dots \beta_p$ are uniquely estimated by
\begin{equation}\label{eq:bhat}
\boldsymbol{\hat\beta} = \left( \mathbf{X}'\mathbf{X}\right)^{-1} \mathbf{X}'\mathbf{Y}
\end{equation}
\subsection{Regularised regression}
Ordinary least squares (OLS) estimates of $\boldsymbol\beta$ are often not appropriate in the analysis of genetic data.  Where predictors are highly correlated, as is often the case for genetic data, OLS estimates $\boldsymbol{\hat\beta}$ can be unstable, having large variance. In a GWAS, nearby SNPs can be exactly correlated, leading to exact collinearity in $\mathbf{X}$. Then, estimates of $\boldsymbol\beta$ are not uniquely defined because the matrix $\mathbf{X}'\mathbf{X}$ is singular and cannot be inverted. When the number of predictors, $p$, exceeds the number of observations, $n$, again OLS estimates of $\boldsymbol \beta$ are not uniquely defined. A number of penalised regression approaches have been proposed to address this problem. Two of these are described here: ridge regression (RR) and principal components regression (PCR).

In both RR and PCR, as in ordinary least squares regression (OLSR), the fitted values of $\mathbf{Y}$ can be expressed as
\begin{equation}\label{eq:yhat}
\mathbf{\hat Y} = \mathbf{H}\mathbf{Y}
\end{equation}
Here, $\mathbf{H}$ is the `hat matrix', or projection matrix, and it relates the fitted values $\mathbf{\hat Y}$ to the observed values $\mathbf{Y}$. In OLSR, RR and PCR, $\mathbf{H}$ is of the form
\begin{equation}
\mathbf{H} = \mathbf{X} \mathbf{G} \mathbf{X'}
\end{equation}
The definition of $\mathbf{G}$ depends on the model being fitted. In OLSR, $\mathbf{G} = \left( \mathbf{X}'\mathbf{X}\right)^{-1}$, (see Equation \eqref{eq:bhat}) and $\mathbf{\hat Y} = \mathbf{X} \left( \mathbf{X}'\mathbf{X}\right)^{-1} \mathbf{X}'\mathbf{Y} = \mathbf{X} \boldsymbol{\hat \beta}$. Both RR and PCR use $\mathbf{G}$ that approximates $\left( \mathbf{X}'\mathbf{X}\right)^{-1}$ in a different way [\cite{Brown93}].

In RR, $\mathbf{G} = \left( \mathbf{X}'\mathbf{X} + k\mathbf{I}_{p} \right)^{-1}$ where $k$ is the ridge parameter and $\mathbf{I}_{p}$ is the $p$-dimensional identity matrix. 

In PCR, $\mathbf{G}$ is formed by taking the eigendecomposition of $\mathbf{X}'\mathbf{X}$:
\begin{equation}\label{eq:eigen}
\mathbf{X}'\mathbf{X} = \mathbf{Q} \boldsymbol{\Lambda} \mathbf{Q}'
\end{equation}
Columns $q_j$ of $\mathbf{Q}$ form the eigenvectors of $\mathbf{X}'\mathbf{X}$ and $\boldsymbol \Lambda = \text{diag}\left( \lambda_1 \geq \lambda_2 \geq  \dots \geq \lambda_{p-1} \geq \lambda_p \right)$ is the diagonal matrix with diagonal elements the eigenvalues of $\mathbf{X}'\mathbf{X}$ in descending order. Then, $\mathbf{G}$ for a PCR using the first $r$ components is given by
\begin{equation}
\mathbf{G} = \sum_{j = 1} ^ r \left( 1 / \lambda_j \right) q_j q_j'
\end{equation}
\subsection{Ridge regression estimators}
As discussed in Section \ref{sec:intro:choice}, a number of methods have been proposed in the literature to estimate $k$ based on the data. One popular estimator, proposed by \cite{Hoerl1975}, is easier to understand when the linear regression model in Equation \eqref{eq:linmodmat} is written in canonical form: 

With columns of $\mathbf{X}$ centred and scaled such that $\mathbf{X}'\mathbf{X}$ is in correlation form, the eigendecomposition of $\mathbf{X}'\mathbf{X}$ (Equation \eqref{eq:eigen}) yields a matrix $\mathbf{Q}$ with columns $q_j$ the eigenvectors of $\mathbf{X}'\mathbf{X}$ and the diagonal elements $\lambda_j$ of $\mathbf{\Lambda}$ the corresponding eigenvalues. 

Taking $\mathbf{Z} = \mathbf{X}\mathbf{Q}$ and $\boldsymbol\alpha = \mathbf{Q}'\boldsymbol\beta$, the model in Equation \eqref{eq:linmodmat} can be rewritten in canonical form as
\begin{equation}\label{eq:linmodcanform}
\mathbf{Y} = \mathbf{Z}\boldsymbol{\alpha} + \boldsymbol{\epsilon}
\end{equation}
Columns of $\mathbf{Z}$ are the $t = \text{min}(n,p)$ principal components of $\mathbf{X}$. The OLS estimator for $\boldsymbol\alpha$ is
\begin{equation}\label{eq:ahat}
\boldsymbol{\hat\alpha} = \mathbf{\Lambda}^{-1}\mathbf{Z}'\mathbf{Y}
\end{equation}
and this is related to $\boldsymbol{\hat\beta}$ as
\begin{equation}\label{eq:ahattobhat}
\boldsymbol{\hat\alpha} = \mathbf{Q}'\boldsymbol{\hat\beta}
\end{equation}
In PCR, $\boldsymbol{\hat\alpha}$ are the regression coefficients of a subset of the PCs that form the columns of $\mathbf{Z}$. In RR, all coefficients $\alpha_i$ are used. With ridge parameter $k$, estimates of RR coefficients are given, in canonical form, by
\begin{equation}\label{eq:ak}
\hat\alpha_k{_j} = \frac{\lambda_j}{\lambda_j + k}\hat\alpha_j 
\end{equation}
These estimated coefficients can be returned to the original form of the model using the transformation in Equation \eqref{eq:ahattobhat}.

\cite{Hoerl1975} propose estimating $\hat k$ from the data as: \begin{equation}\label{eq:kHKB}
k_\text{HKB} = \frac{p\hat\sigma^2}{\boldsymbol{\hat\alpha} ' \boldsymbol{\hat\alpha}} =  \frac{p\hat\sigma^2}{\boldsymbol{\hat\beta} ' \boldsymbol{\hat\beta}} 
\end{equation}

This ridge estimator is motivated as the harmonic mean of the `ideal' generalized ridge estimator in terms of minimizging MSE. $\boldsymbol{\hat\alpha}$ in this estimator are the PCR coefficients, and are estimated as in equation \eqref{eq:ahat}. $\hat\sigma^2$ is estimated by 
\begin{equation}\label{eq:sigma2}
\hat\sigma^2 = \frac{(\mathbf{Y} - \mathbf{X}\boldsymbol{\hat\beta})'(\mathbf{Y} - \mathbf{X}\boldsymbol{\hat\beta})}{n - p} 
\end{equation}

We also investigated the performance of ridge estimators based on that proposed by \cite{Lawless:1976p100}. However we found that these performed less well than estimators based on the estimator of \cite{Hoerl1975}.

As mentioned in Section \ref{sec:intro:choice}, the ridge estimator $k_\text{HKB}$ (Equation \eqref{eq:kHKB}) is not useful when the data comprise more predictors than observations, as in such situations OLS estimates of the parameters $\sigma^2$ and $\boldsymbol\beta$ do not exist, resulting in $k_\text{HKB}$ being undefined. However, with the model in canonical form, $k_\text{HKB}$ is written as:
\begin{equation}
k_\text{HKB} = \frac{p \hat \sigma^2} {\boldsymbol{\hat\alpha}'\boldsymbol{\hat\alpha}}
\end{equation}
\begin{equation}\label{eq:sig2hat}
\hat\sigma^2 = \frac{(\mathbf{Y} - \mathbf{Z}\boldsymbol{\hat\alpha})'(\mathbf{Y} - \mathbf{Z}\boldsymbol{\hat\alpha})}{n - p} 
\end{equation}
We consider a regression situation in which most of the variance in $\mathbf{X}$ is explained by its first few PCs. Then, we propose a ridge penalty that uses the harmonic mean of the `ideal' penalties of only these first $r$ components, resulting in a ridge penalty of the form:
\begin{equation}\label{eq:kr}
k_r = \frac{r\hat\sigma_r^2}{\sum_{j=1}^r\hat\alpha_j^2}
\end{equation}
Here, 
\begin{equation}\label{eq:sig2hatr}
\hat\sigma^2_r = \frac{(\mathbf{y} - \mathbf{Z_r}\mathbf{\hat{\alpha}_r})'(\mathbf{y} - \mathbf{Z_r}\mathbf{\hat{\alpha}_r})}{n - r} 
\end{equation}
and $\mathbf{Z_r}$ and $\mathbf{\hat\alpha_r}$ contain the first $r$ columns and the first $r$ elements of $\mathbf{Z}$ and $\boldsymbol{\hat\alpha}$ respectively. Before $k_r$ can be used in practise, we must decide the number of PCs, $r$, to use in its calculation.

\begin{figure}[t] 
  \includegraphics[angle = 270, width = \textwidth]{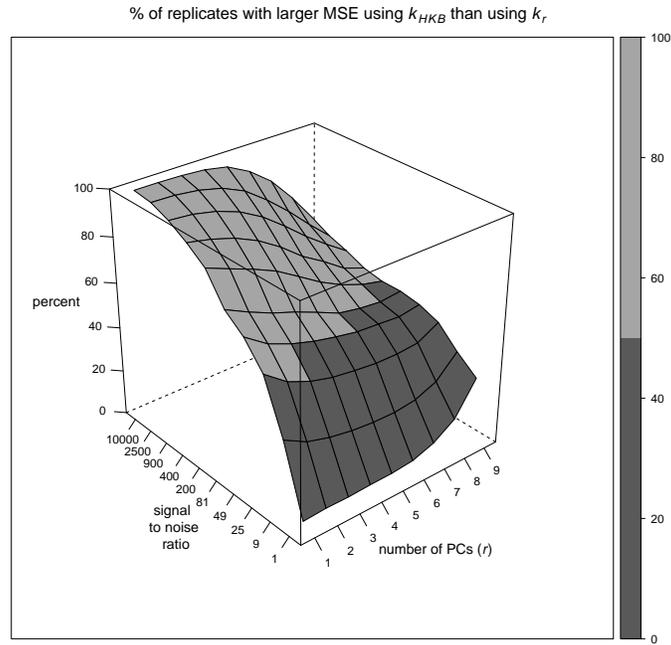}
\vspace{6pc}
\caption[]{Comparing the mean-squared error of ridge regression estimates obtained using the shrinkage parameter $k_r$ to those obtained using the shrinkage parameter $k_\text{HKB}$. Plotted are the proportion of replicates that using $k_r$ results in \emph{larger} mean squared error than the estimates fitted using $k_\text{HKB}$ (equivalent to $k_r$ with $r = p$).}
\label{fig:KrVsKhkb}
\end{figure}

As a starting point, we address whether it is more useful to include all non-zero PCs (of which there are at most $\text{min}(n, p)$), or to include fewer than all the non-zero PCs. To this end we reanalyse the data analysed by \cite{Hoerl1975}, extending their results to compare the shrinkage parameter $k_r$ to $k_\text{HKB}$. The data being reanalysed are a ten-factor dataset consisting of 36 observations. These data were first discussed by \citet{Gorman:1966p1476} and are described in \citet{Daniel:1999p1810}. They relate to describing the operation of a petroleum refining unit. Following the approach taken by \cite{Hoerl1975}, we use the ten-factor dataset as a design matrix in a simulation study. In each replicate, a vector of regression coefficients with a specified squared length is simulated. As in \cite{Hoerl1975} we find that, subject to normalisation, our results are not sensitive to this value. Response variables are simulated at a range of signal-to-noise ratios. For each signal-to-noise ratio, 1000 replicates are simulated, and results are reported as an average of these. \cite{Hoerl1975} tabulate the mean squared error under both the linear and ridge models and report the percentage of replicates linear regression gives rise to estimates $\boldsymbol{\hat\beta}$ with smaller smaller mean squared error than ridge estimates $\boldsymbol{\hat\beta}_{k_\text{HKB}}$ with $k_\text{HKB}$ defined as in Equation \eqref{eq:kHKB}. Following this approach, in Figure \ref{fig:KrVsKhkb} we plot the percentage of replicates that $k_r$ results in ridge estimates $\boldsymbol{\hat\beta}_{k_r}$ with smaller mean squared error than the estimates obtained using the shrinkage parameter $k_\text{HKB}$. From this figure we see that, when the signal to noise ratio is not too low, estimates of $\boldsymbol{\hat\beta}$ with smaller mean squared error are obtained using $k_r$ with $r < p$ than when using $k_\text{HKB}$.

With evidence that using $k_r$ with  $r < p$ as a shrinkage estimator can result in improved estimates of $\boldsymbol{\hat\beta}$ compared to when $k_\text{HKB}$ is used, our method is naturally extensible to data with more predictors than observations ($p>n$). Whilst OLS estimators are not defined when there are more predictors than observations, the eigenvectors and corresponding eigenvalues of the correlation matrix can be determined, and the first $r$ of these used to compute $k_r$. Again, we need to determine whether inclusion of all nonzero PCs results in estimates with smallest prediction error or whether smaller prediction error is obtained when using the first $r$ PCs, $r < n$. In Table \ref{Tab:PSEvsPropVar} we report average PSE when $k_r$ is calculated using different numbers of PCs, in simulated genetic data with more predictors than observations. Because results are averaged over ten replicates, instead of reporting PSE at different values of $r$, we report PSE at different cumulative proportions of variance explained by the the PCs, because in different replicates the correlation structure, and thus the proportion of variance explained by a given number of principal components, differs. We see that minimum PSE is obtained when fewer than the maximum number of PCs are used to compute $k_r$. 
\subsection{Semi-automatic choice of $k$}\label{sec:biasvar}

We investigate two methods for choosing the number of PCs, $r$, to use in the computation of $k_r$. We use the cross-validated PRESS statistic [\cite{Balding:2007handbook}], and we use the number of PCs such that the degrees of freedom for variance of the model is equal to the number of PCs used in its calculation. This results in a model with the same degrees of freedom for variance as a PCR using $r$ PCs, but with the penalization shared out among all the PCs. 


We are interested in the predictive performance of a model on independent test data. For a regression fit which can be expressed in the form of Equation \eqref{eq:yhat}, PSE can be written as:
\begin{equation}
\text{PSE} = \sigma^2 + \frac{\text{tr}\left(\mathbf{H} \mathbf{H}'\right)}{n} \sigma^2 + \frac{\mathbf{b}'\mathbf{b}}{n}
\end{equation}
where $\mathbf{b} = \mathbf{Y} - \mathbf{X}\boldsymbol{\hat\beta}$ is the bias, the distance between the fitted estimates and the true ones. The first term measures the (unavoidable) noise in $\mathbf{Y}$, the second measures variance in the prediction estimates. Above, we have determined expressions for $\mathbf{H}$ in OLSR, RR and PCR. In OLSR, $\text{tr}\left( \mathbf{H}\mathbf{H}'\right) = p$, the number of parameters in the regression fit. The aim in penalised regression is to reduce the variance whilst introducing a little bias, keeping the overall $\text{PSE}$ lower than that of OLSR. In PCR, $\text{tr} \left( \mathbf{H} \mathbf{H}' \right) = r$, the number of components used in the penalised regression fit. 

In RR, it is straightforward to find a shrinkage parameter $k$ such that $\text{tr} \left( \mathbf{H}_k\mathbf{H}_k'\right) = r $ where $r$ is any specified value, by noting that $\text{tr}\left( \mathbf{H}_k \mathbf{H}_k' \right) = \sum_{j = 1}^p \frac{\lambda_j^2}{\left( \lambda_j + k \right)^2}$ and using numerical methods to find $k$. Thus, we can compare PCR and RR in terms of prediction error when the variance, $\frac{\text{tr}(\mathbf{H} \mathbf{H}') }{n} \sigma^2$, is equal in both models. With $\text{Var}( \hat Y )=\sigma^2 $ fixed, we are only interested in the bias term and we can find an expression for this also, by noting that:
 
\begin{equation}\label{eq:bias}
\mathbf{b} = \left[\mathbf{X} - \mathbf{X}  \mathbf{G}          \mathbf{X}' \mathbf{X}\right]\boldsymbol\beta 
\end{equation} 

 In OLSR, the estimates are unbiased ($\mathbf{b} = 0$).
In a simulation study with $\boldsymbol \beta$ known, we can compare the bias in RR and PCR when the variance of the fitted $\mathbf{\hat Y}$ in each model is fixed such that $\text{tr}\left( \mathbf{H} \mathbf{H}' \right) = r$. 

In PCR, the coefficients of the first $r$ PCs are their least squares counterparts; the coefficients of the remaining components are set to zero. Thus the bias is the difference between zero and the least squares estimate of the coefficient of the $r+1 \dots t^\text{th}$ components, where $t = \text{min}(n,p)$ is the maximum number of PCs.

In RR the bias is more `spread out' among the $t$ components as the least squares estimate of each coefficient is `shrunk' by $\frac{\lambda_j}{\lambda_j + k}$.

We illustrate this using a simulation study. The patterns of predictors and coefficients used by \cite{Zou:2005p598} are used here, to cover a range of parameter values and correlation structures. All scenarios comprise more observations than predictors; nonetheless they illustrate the bias-variance decomposition of the PSE. The error in $\mathbf{Y}$ is accounted for by other terms in the prediction error; therefore in calculating the bias we consider only the predictors and coefficients. The four regression scenarios are detailed in Table \ref{tab:ZouHastie}. 

\begin{figure}[t] 
 \begin{center}
 \includegraphics[width=0.4\textwidth, angle=270]{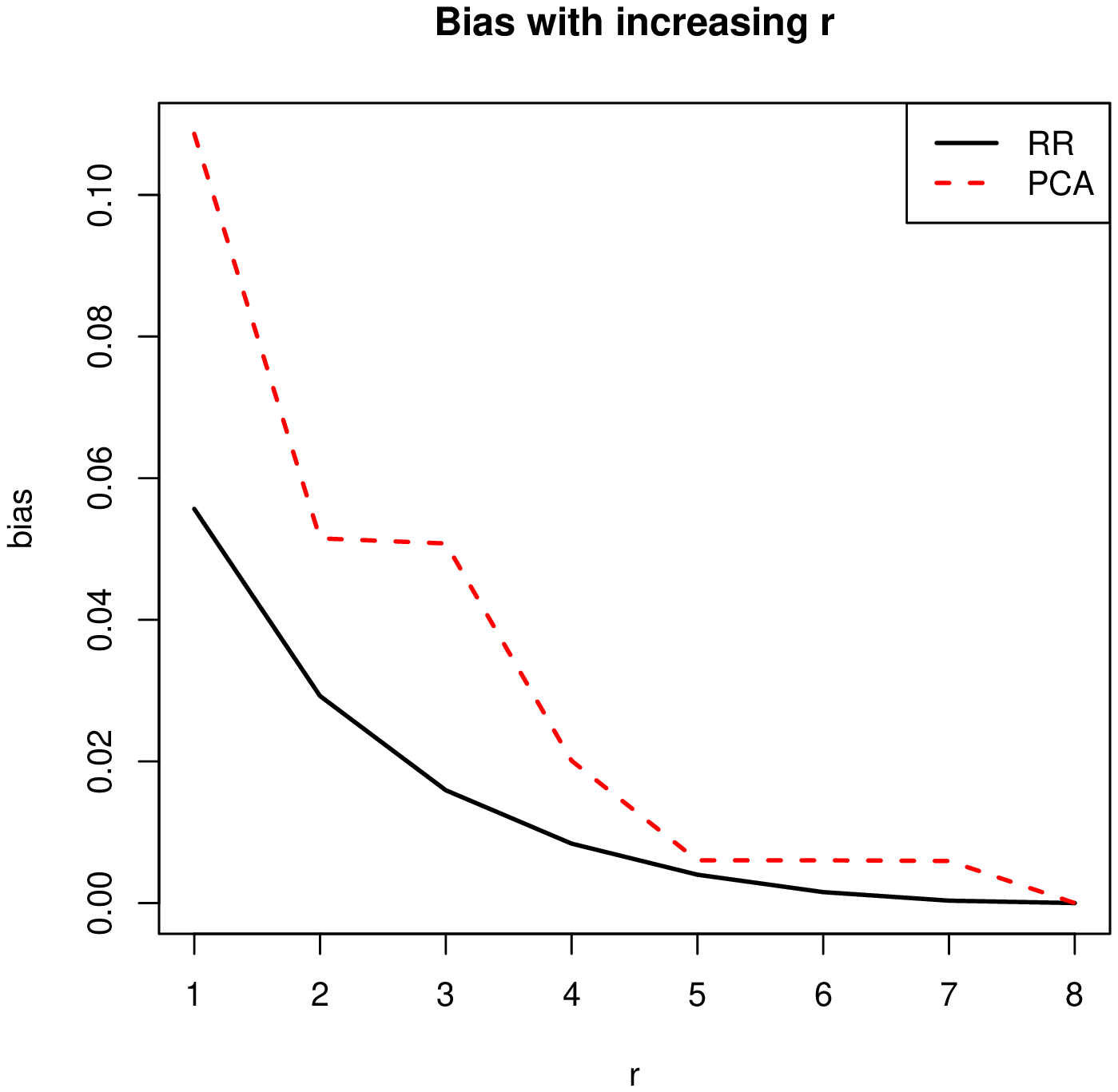}
 \includegraphics[width=0.4\textwidth, angle=270]{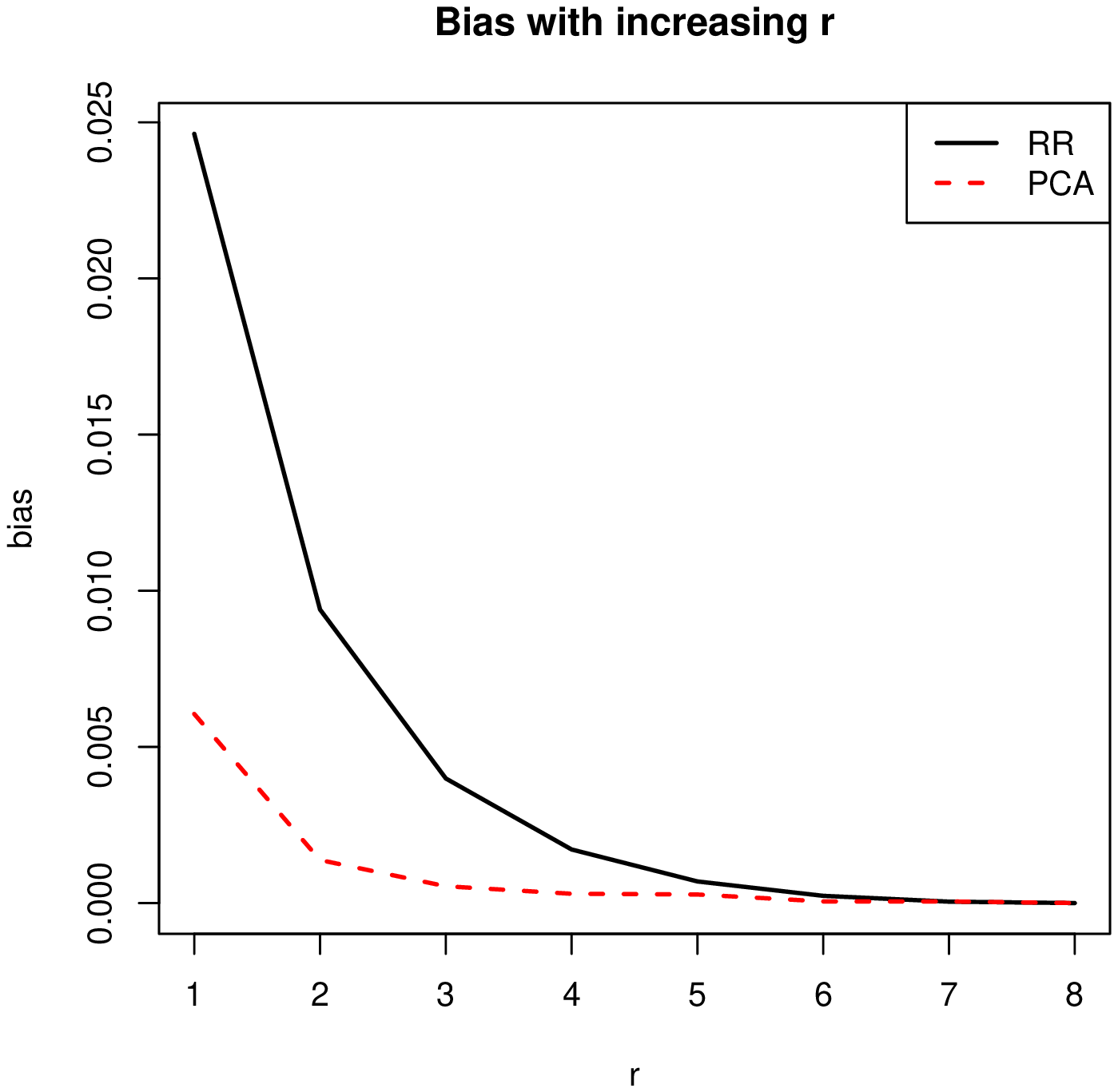} \\
 \includegraphics[width=0.4\textwidth, angle=270]{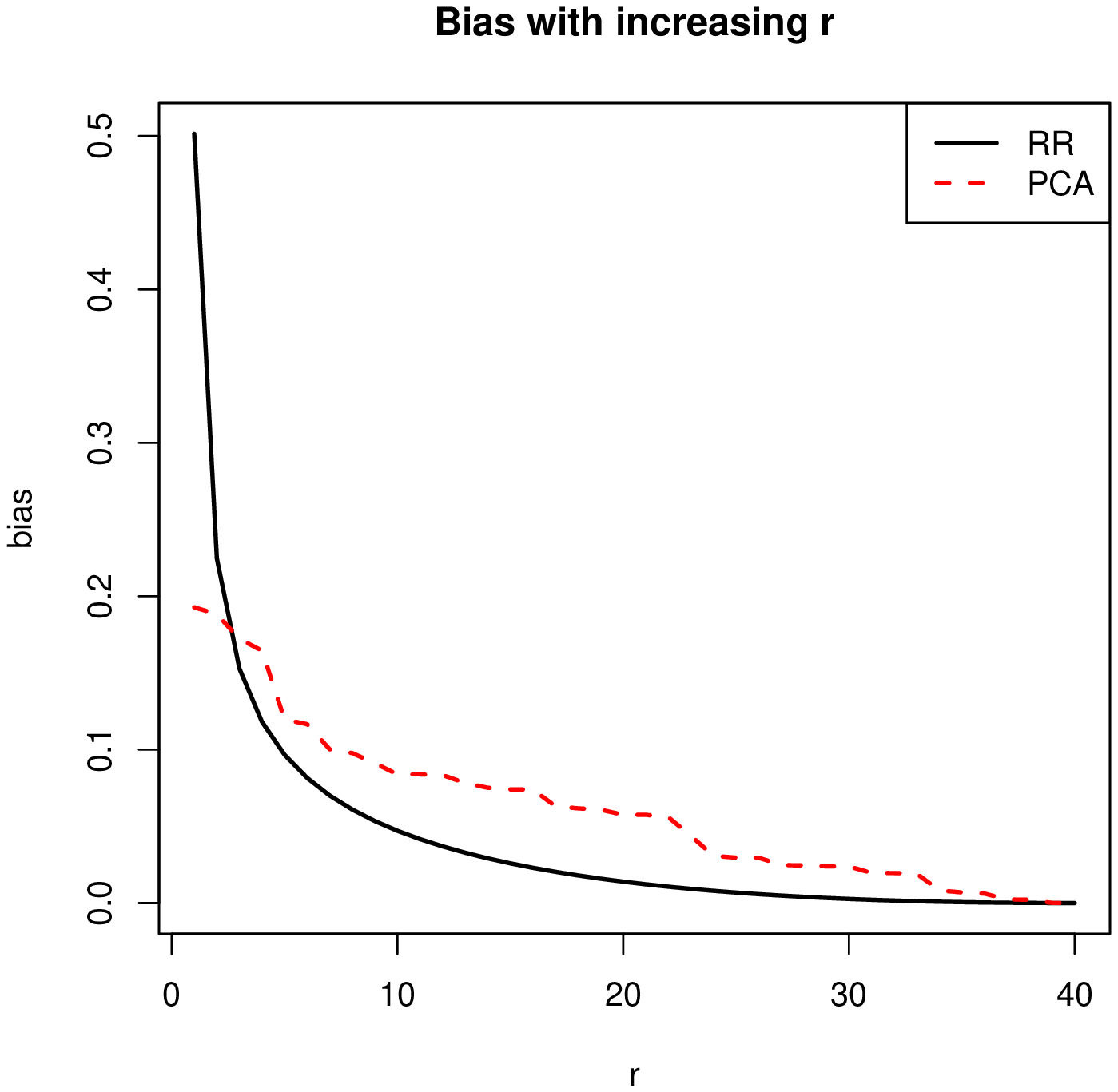}
 \includegraphics[width=0.4\textwidth, angle=270]{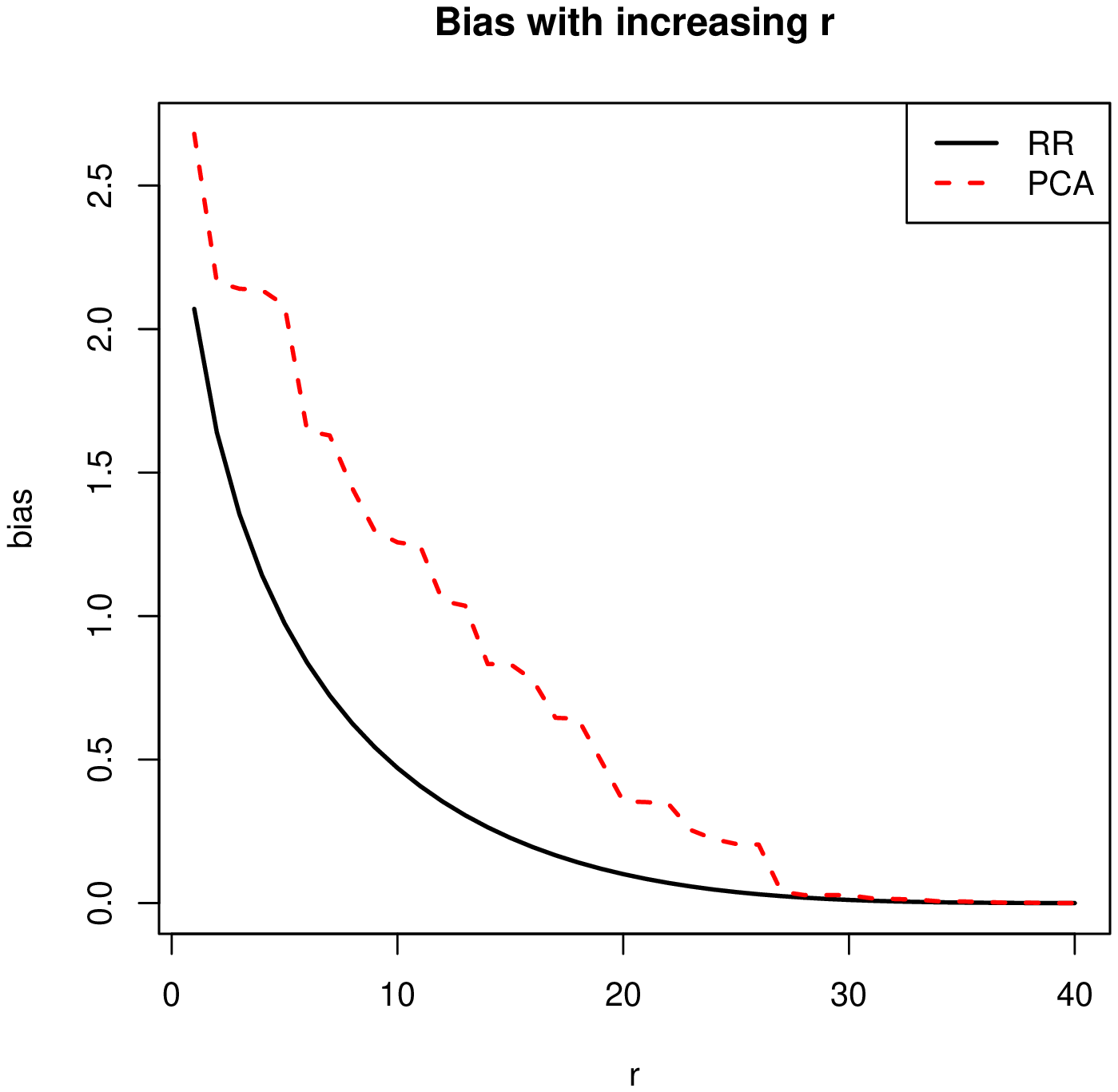}
\caption[]{Bias $\left(\frac{\mathbf{b}'\mathbf{b}}{n}\right)$ in PCR and RR in regression scenarios (1), (2), (3) and (4) (Table \ref{tab:ZouHastie}), at different values of $r$.}
\label{penG}
 \end{center}
\end{figure}

In Figure \ref{penG}, we plot $\mathbf{b}'\mathbf{b}/n$ using $\mathbf{b}$ defined as in Equation \eqref{eq:bias} for $r = 1, \dots, t$. We see that for regression scenarios (1), (3) and (4), the bias is typically lower for RR than for PCA. The only scenario in which PCA has lower bias than RR is scenario (2), where there is moderate correlation among the predictors but all the coefficients have the same effect size, a situation that is unlikely to arise in genetic data. We can see the smooth decrease in the bias with RR whereas in PCR the bias decreases in a stepwise manner, with each step corresponding to the inclusion of one more component in the model. As $r$ approaches $t$, the fitted coefficients approach their least squares counterparts and the bias approaches 0, its value in OLSR. 
\subsection{Degrees of Freedom: other definitions}\label{sec:doff}
OLSR, RR and PCR all result in models of the form given in Equation \eqref{eq:yhat}. For models that can be expressed in this form, several definitions of effective degrees of freedom have been proposed [\cite{Hastie1990}].

The effective number of parameters, $\text{tr}(\mathbf{H})$, gives an indication of the amount of fitting that $\mathbf{H}$ does. As discussed in section \ref{sec:biasvar}, $\text{tr}(\mathbf{H}\mathbf{H}')$ can be defined as the effective degrees of freedom for variance. The degrees of freedom for error, as used in the denominator of the estimate of $\sigma^2$ (Equation \eqref{eq:sigma2}) is given by $n - \text{tr}(2\mathbf{H} - \mathbf{H}\mathbf{H}')$, thus the effective number of parameters in the degrees of freedom for error is $\text{tr}(2\mathbf{H} - \mathbf{H}\mathbf{H}')$.

In OLSR, RR and PCR it can be shown that $\text{tr}(\mathbf{H}\mathbf{H}') \leq \text{tr}(\mathbf{H}) \leq \text{tr}(2\mathbf{H} - \mathbf{H}\mathbf{H}')$ (\cite{Hastie1990}, see Appendix \ref{DegOfFAppendix}). In OLSR, all three definitions of degrees of freedom reduce to to $p$, the number of parameters in the model. In PCR, all three definitions reduce to $r$, the number of components retained in the PCR. In RR with $k > 0$, the three definitions take values that follow the above inequalities. For each of the definitions, it is possible to choose the ridge parameter such that the effective degrees of freedom equal some specified value. For a fixed value of the effective degrees of freedom, it can be shown that $k_{\mathbf{H}\mathbf{H}'} < k_\mathbf{H} < k_{2\mathbf{H} -\mathbf{H}\mathbf{H}'}$ where $k_{\mathbf{H}\mathbf{H}'}$ is $k$ such that $\text{tr}(\mathbf{H}\mathbf{H}') = r$, $k_\mathbf{H}$ is $k$ such that $\text{tr}(\mathbf{H}) = r$ and $k_{2\mathbf{H} - \mathbf{H}\mathbf{H'}}$ is $k$ such that $\text{tr}(2\mathbf{H} - \mathbf{H}\mathbf{H}') = r$ (for the same value of $r$ in all three cases). The larger the value of $k$, the further the ridge estimates are from the OLS estimates. Thus, choosing $k$ such that $\text{tr}(\mathbf{H}\mathbf{H}') = r$ (among the three definitions of degrees of freedom) results in regression coefficient estimates that are closest to the OLS estimates. The proof of these assertions is given in Appendix \ref{DegOfFAppendix}.

\subsection{A prior on the number of components}
RR has a Bayesian interpretation. RR coefficients are the maximum {\em a posteriori} estimates in a regression setting where, using the canonical form of the linear regression model: 

\begin{equation}
\boldsymbol{Y} \sim \mathcal{N} \left( \mathbf{Z}\boldsymbol{\alpha} , \sigma^2 \mathbf{I} \right)
\end{equation}

Further, we assume that elements of $\boldsymbol\alpha$ are exchangeable and have a prior distribution
\begin{equation}
\alpha_j \sim \mathcal{N} \left( 0, \sigma^2_\alpha \right)
\end{equation}

Then, a Bayes estimator for $\boldsymbol\alpha$ is $\boldsymbol\alpha^*$, [\cite{Lawless:1976p100}] with
\begin{equation}
\hat\alpha^*_j = \frac{\lambda_j}{\lambda_j + \frac{\sigma^2}{\sigma^2_\alpha}}\hat\alpha_j 
\end{equation}

If we estimate $\sigma^2$ by $\hat\sigma^2$ as in Equation \eqref{eq:sig2hat}, and set $\sigma^2_\alpha$ equal to $ \frac{  \boldsymbol{\hat\alpha} '  \boldsymbol{\hat\alpha}}{p}$, then we obtain the ridge estimator $k_\text{HKB}$ (Equation \eqref{eq:kHKB}) of \cite{Hoerl1975}. To obtain the proposed estimator $k_r$ (Equation \eqref{eq:kr}), we estimate $\sigma^2$ by $\hat\sigma^2_r$ (Equation \eqref{eq:sig2hatr}), and set $\sigma^2_\alpha$ equal to $ \frac{ \sum_{j = 1}^r \hat\alpha_j^2 }{r}$.

To continue with the estimation of $\boldsymbol\alpha$ from a Bayesian perspective, we can put a prior on the number of components, $r$. Then, the marginal prior on $\boldsymbol\alpha$ becomes 
\begin{equation}\label{eq:marginalPrior}
\boldsymbol{\alpha} | \boldsymbol{X}, \boldsymbol{Y} \sim  \mathcal{N} \left( 0,  \sum_{r=1}^p \text{Pr}\left( R = r \right)   \sigma^2_{\alpha_r} \right)
\end{equation}

We considered the effect of different prior distributions on $r$ on the marginal prior on $\boldsymbol\alpha$. Simulation studies showed that the use of the proposed estimator $k_r$ with $r$ chosen such that $r = \text{tr}\left( \mathbf{H}'\mathbf{H}\right)$ results in a ridge regression estimator that shrinks the coefficients more than when a uniform prior on $r$, $\text{Pr}\left(R = r\right) = \frac{1}{\text{min}(n,p)}$, was used in Equation \eqref{eq:marginalPrior} (results not shown). 

\subsection{Ridge Logistic Regression}
The logistic regression model is commonly used to model the effect of one or more predictors when the response is binary. In logistic regression, similar problems arise when estimating regression coefficients when data are highly correlated or high dimensional. Maximum likelihood estimates may have large variance or, in cases of exact collinearity or more predictors than observations, be undefined. Ridge logistic regression has been considered in the literature as a means of addressing these difficulties. \cite{Schaefer:1984p51} propose a `Ridge type' estimator and demonstrate that it will result in coefficient estimates with smaller mean squared error than the maximum likelihood estimator when the predictor variables are collinear. The ridge penalty proposed by \cite{Schaefer:1984p51} is:
\begin{equation}
k = \frac{p}{\boldsymbol{\hat\beta}'{\boldsymbol{\hat\beta}}}
\end{equation}
\cite{Lee:1988p928} consider this ridge estimator together with an alternative, and find the performance in terms of mean squared error depends on the structure of the data. \cite{Cessie1992} use generalized cross-validation to guide their choice of ridge parameter.

Principal components logistic regression (PCLR, \cite{Aguilera:2006p2476}) has been used to overcome the problems associated with correlated predictors in logistic regression. In PCLR, the PCs of $\mathbf{X}$ are used as covariates in the logistic regression model. We use increasing numbers of PCs in a PCLR to compute the shrinkage parameter and degrees of freedom. For a PCLR using $r$ components, the corresponding penalty is calculated as
\[
k_\mathbf{r} = \frac{r}{ \boldsymbol{\hat{\alpha}}_r' \boldsymbol{\hat{\alpha}}_r }
\]
where $\boldsymbol{\hat{\alpha}}_r$ is the vector of $r$ regression coefficients computed using PCLR.

To estimate the effective degrees of freedom of the logistic regression model, we used the trace of the square of the hat matrix (as in linear regression). The hat matrix is calculated as:
\[
\mathbf{H}=\left( \mathbf{X}'\mathbf{W}\mathbf{X} + k\mathbf{I}\right)^{-1}\mathbf{X}'\mathbf{W}\mathbf{X}
\]
where $\mathbf{W} = \text{diag}(\hat\pi(\mathbf{x}_i)(1 - \hat\pi(\mathbf{x}_i)))$. $\hat\pi(\mathbf{x}_i)$ are the fitted probabilities, see Appendix \ref{CLG}. 
As in linear RR, we compute the effective degrees of freedom for variance as $\text{tr}\left( \mathbf{H}\mathbf{H}'\right)$.
We use simulations studies to evaluate several guidelines for choosing the number of components $r$ to use in the calculation of $k_r$. We apply these guidelines to a real data example. 

\section{Simulation study}
\label{sec:simstudy}
To evaluate the properties of the proposed shrinkage estimator, $k_r$, we performed simulation studies.

Firstly, we sought a rule for the number of PCs, $r$, to use to compute $k_r$. Because the aim was to identify a shrinkage estimator with good predictive performance, we compared the predictive performance of RR models fitted using $k_r$ when $r$ took a range of values. We also considered two rules that determined how many PCs to use: choosing $r$ such that the degrees of freedom for variance, $\text{tr}(\mathbf{H}\mathbf{H}')$, is equal to $r$ (see section \ref{sec:doff}), and determining the best value of $r$ using the cross-validated PRESS statistic [\cite{Balding:2007handbook}]. The cross-validated PRESS statistic selects the number of PCs that minimize the in-sample prediction error.

In the second part of our simulation studies, we used the rule from the first part that had best predictive performance. We then compared the predictive performance of RR models fitted using $k_r$ to that of competing regression approaches that can be applied to high-dimensional data.

\subsection{Simulated data}

Because our motivation is risk prediction using genetic data, we used simulated genetic data for the simulation studies. The data were simulated SNP data generated using the software FREGENE [\cite{Hoggart:2007p111, ChadeauHyam:2008p132}] as a panmictic population of 21,000 haplotypes. The simulated data are available for download from \url{http://www.ebi.ac.uk/projects/BARGEN/}. We used a region of approximately 7Mb, containing 20,000 SNPs with minor allele frequency (MAF) $>1\%$. 

Genotype and corresponding phenotype data were simulated with both continuous and binary outcomes, analysed using linear and logistic regression respectively. Each replicate consisted of 1,000 training individuals and 500 test individuals, and results were averaged over ten replicates. Data were generated and analysed as follows:

\begin{description}
	\item[Continuous outcomes] - analysed using linear regression. 200 SNPs with MAF 10 - 15 \% were randomly selected to be causal SNPs; these causal SNPs were assigned an effect size drawn from a  uniform distribution $U [0.05, 0.1]$. All other SNPs were given an effect size of 0. Thus the vector of effect sizes, $\boldsymbol\beta$, of length 20,000, contained 200 non-zero elements. Genotypes were generated by summing two randomly selected haplotypes. Responses were generated as $\mathbf{Y} = \mathbf{X}\boldsymbol\beta + \epsilon, \epsilon \myiid \mathcal{N}(0, 1)$. Model performance was evaluated using prediction squared error (PSE) over the test data:
	\begin{equation}
		\text{PSE} = \frac{1}{n}\sum_{i=1}^n \left( Y_i - \hat{Y}_i\right)^2
	\end{equation}
	where $\hat{Y_i}$ is the fitted outcome of the $i^{\text{th}}$ individual.
	\item[Binary outcomes] - analysed using the logistic model. Case-control outcomes were generated by randomly selecting two haplotypes which were summed to generate a simulated genotype. The probability of an individual with that genotype being a case was generated as $\text{Pr}\left(Y_i = 1 | \mathbf{x}_i\right) = \text{exp}(-5 + \mathbf{x}_i'\boldsymbol\beta) / (1 + \text{exp}(-5 + \mathbf{x}_i'\boldsymbol\beta))$, and that individual's case-control status was determined randomly according to this probability. This process was repeated until the required (equal) number of cases and controls was obtained.
	
For the data with binary outcomes, predictive performance was measured using classification error, as in \cite{Cessie1992}:
	\begin{equation}
		\text{CE}_i = \left\{ \begin{array}{l l}
    						0 & \quad Y_i =0, \hat{\pi}\left( \mathbf{x}_i\right) < 0.5  || Y_i =1, \hat{\pi}\left( \mathbf{x}_i\right) > 0.5 \\
						\frac{1}{2} & \quad \hat{\pi}\left( \mathbf{x}_i\right) = 0.5 \\
    						1 & \quad Y_i =1, \hat{\pi}\left( \mathbf{x}_i\right) < 0.5  || Y_i =0, \hat{\pi}\left( \mathbf{x}_i\right) > 0.5 \\
  					\end{array}\right. 
	\end{equation}
		Here, $\hat{\pi}\left( \mathbf{x}_i\right)$ is the estimated probability that the $i^\text{th}$ individual is a case based on his genotypes, that is  $\text{pr} \left(Y_i = 1 | \mathbf{x}_i \right)$. In the results we present, we take the average CE:
	\begin{equation}
		\text{Average CE} = \frac{1}{n} \sum_{i = 1}^n \text{CE}_i 
	\end{equation}
\end{description}

\subsection{Simulation study results}

First we considered the predictive performance of $k_r$ using different numbers of PCs. Because PCs explain different proportions of variance in different simulation replicates, instead of comparing results at different values of $r$ we compare predictive performance at different proportions of variance explained. Table \ref{Tab:PSEvsPropVar} shows prediction squared error or classification error at different proportions of variance explained. In the column labelled MAX, $r$ is the maximum number of PCs where the corresponding eigenvalues are non-zero. CV is $k_r$ with $r$ chosen using the cross-validated PRESS statistic, and DofF is $k_r$ chosen such that $\text{tr}(\mathbf{H}\mathbf{H}') = r$. From this table we see that, for both continuous and binary outcomes, the best predictive performance is obtained when somewhat fewer than the maximum number of PCs is used to compute $k_r$. We see that of the two rules we investigated, choosing $r$ such that $\text{tr}(\mathbf{H}\mathbf{H}') = r$ offers best predictive performance.

\begin{figure}[t] 
  \includegraphics[angle = 270, width = 0.49\textwidth]{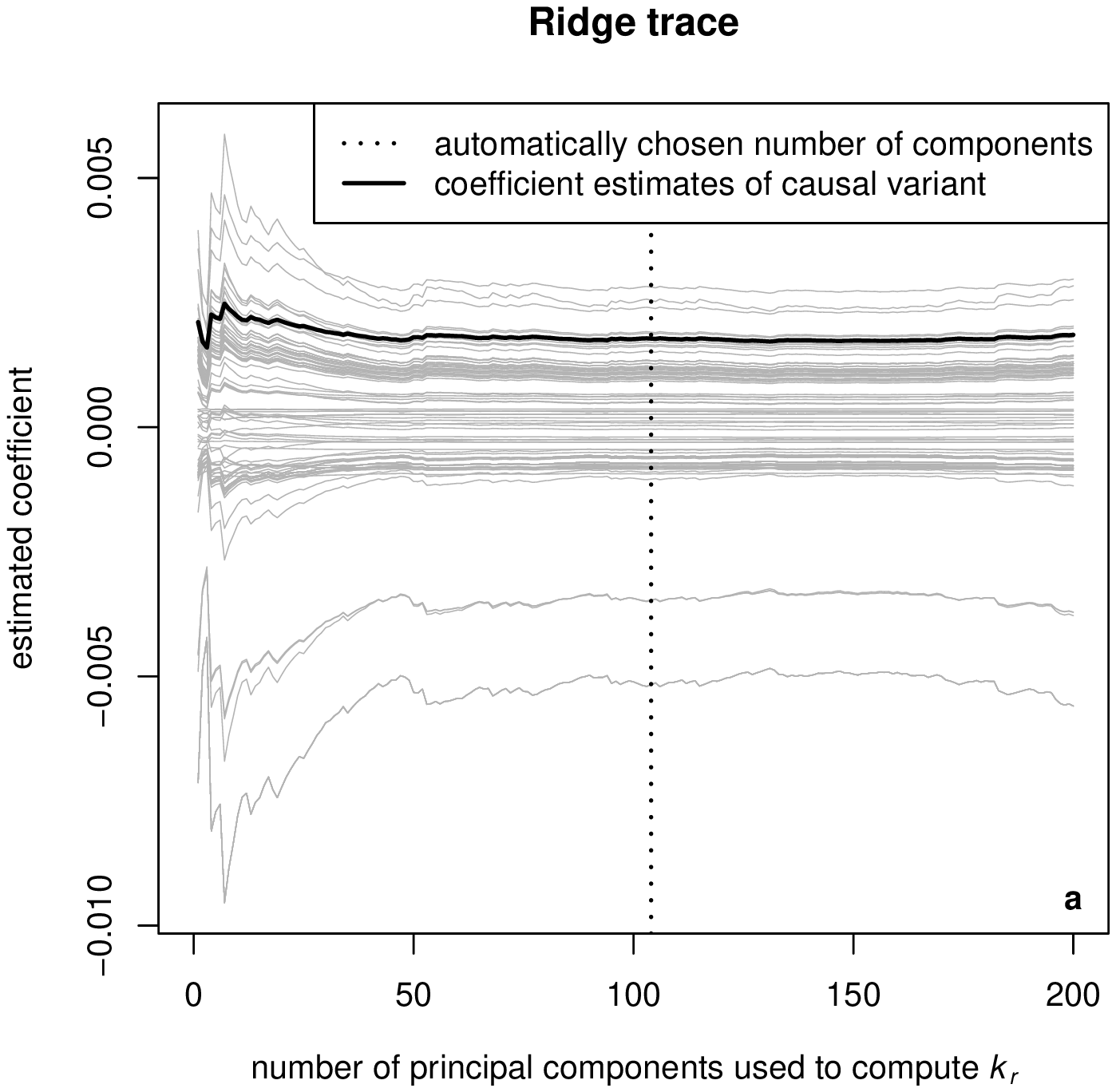}
    \includegraphics[angle = 270, width = 0.49\textwidth]{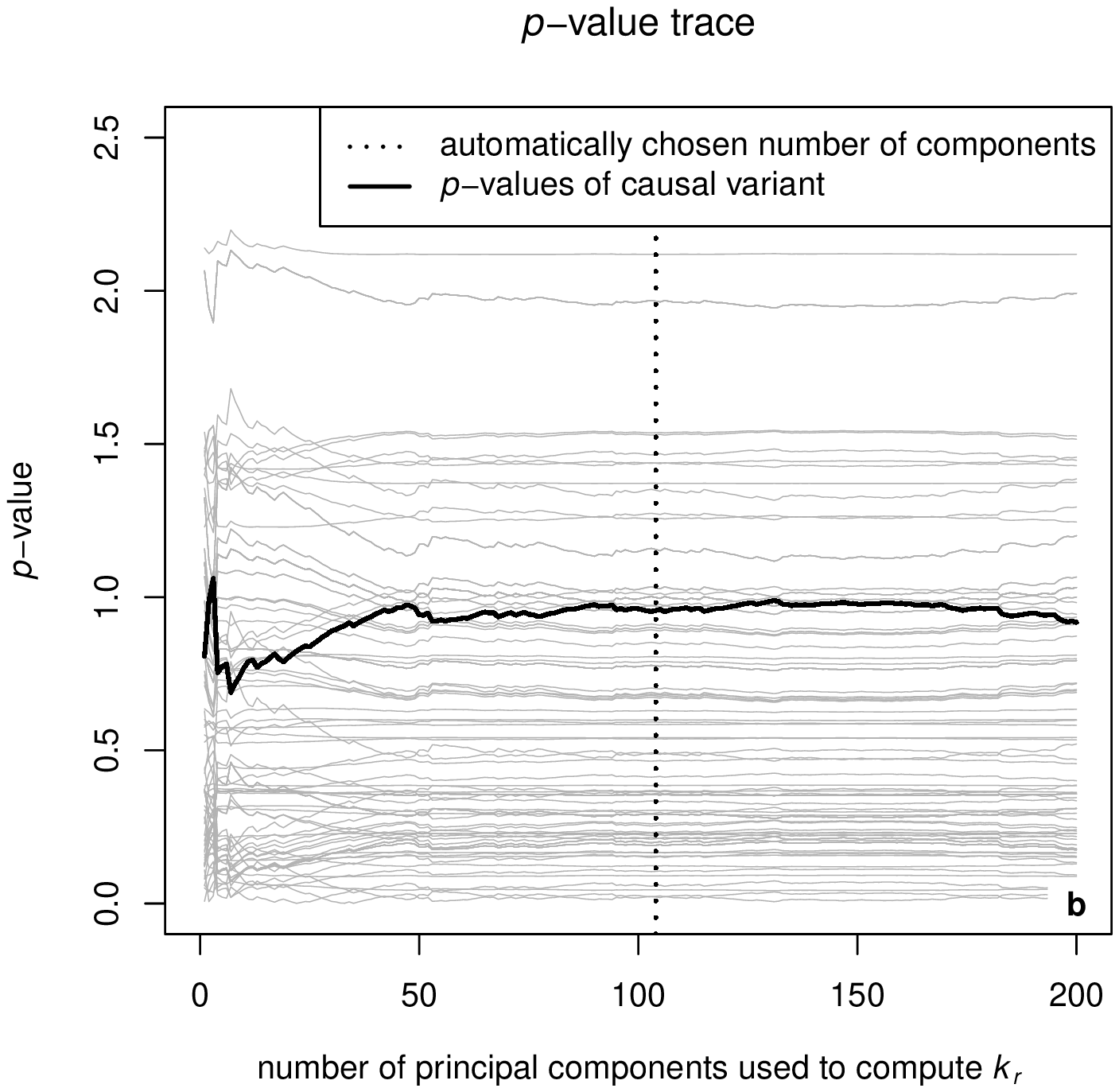}
\vspace{6pc}
\caption[]{{\bf a.} Ridge trace showing estimated regression coefficients and {\bf b.} $p$-value trace of RR coefficients estimated using $k_r$ computed using different numbers of PCs. The $x$-axis shows the number of PCs used to compute $k_r$. The vertical dotted line indicates that our proposed method of choosing the number of components, $r$, chooses a ridge parameter in the region where ridge estimates and their corresponding $p$-values stabilise. The black line indicates a causal variant. Plotted are the first 100 SNPs of the 20,000 in one simulation replicate, with continuous outcomes.}
\label{fig:ridgetrace}
\end{figure}

In Figure \ref{fig:ridgetrace}a we plot the fitted RR coefficients with different values of $r$ used to compute $k_r$; Figure \ref{fig:ridgetrace}b shows the corresponding $p$-values. These plots are taken from one simulation replicate with continuous outcomes. The number of components chosen using $\text{tr}(\mathbf{H}\mathbf{H}') = r$, is indicated and a causal variant is highlighted. We see that using this rule to choose $r$ results in a shrinkage parameter in a region of the ridge trace, and the corresponding $p$-value trace, where the RR coefficients and their corresponding $p$-values are stable and do not change much with further increases in the number of principal components. 

Having demonstrated that the rule of choosing $r$ such that $\text{tr}(\mathbf{H}\mathbf{H}') = r$ performs well among different ways to choose $r$, we compared the performance of RR using $k_r$ to two competing methods of fitting prediction models to high dimensional data. The two other methods were:

\begin{enumerate}
\item Variable selection followed by multivariate regression. Univariate linear or logistic regression was used to estimate the strength of association of each predictor variable with the outcome. A proportion of the predictor variables that were most strongly associated with the outcome in univariate tests were included in a multiple linear or logistic regression model. When more than one predictor was in high LD and both predictors cannot be included in a multiple regression model simultaneously, only one of the correlated predictors was included. Because the number of causal variables is not known $a priori$, we evaluated the predictive performance when a range of proportions of predictors were included in the multiple regression. 
\item HyperLasso regression [\cite{Hoggart:2008p147}] is a penalized regression method that simultaneously considers all predictor variables in a high-dimensional regression problem. HyperLasso was originally applied to the problem of identifying causal variables when predictors are correlated, but it was shown that by using a less stringent threshold for inclusion of predictors in the model, HyperLasso could be used to address the problem of prediction. In order to obtain good predictive performance, it is necessary to relax the penalty so that sufficient coefficients are estimated as non-zero.  The penalty in HyperLasso regression is such that, among a group of correlated predictors, only one coefficient will be estimated as non-zero. This is a disadvantage in prediction using genetic data, where several correlated predictors may contain information that is useful for prediction even if if they are not all causal variables. 

HyperLasso requires the specification of two parameters to control the amount of shrinkage. Following \cite{Eleftherohorinou:2009p2336} the shape parameter in HyperLasso regression was fixed to 3.5 and the penalty parameter was chosen using ten-fold cross-validation. 
\end{enumerate} 

In Table \ref{Tab:SmallEffect} we report the results, comparing mean prediction squared error or mean classification error using these three different regression approaches. We see that RR outperforms the competing methods for this regression problem, which is a realistic simulation of the problem of risk prediction in genetic data. Using univariate variable selection followed by multiple regression, the best performance was obtained when the number of predictors included in the model was equal to the number of non-zero regression coefficients when the data were generated, a proportion that would not be known in practise. In HyperLasso regression, the cross-validation required to choose the penalty parameter was computationally demanding, and we found in order to obtain the best predictive performance a relaxed penalty was necessary. RR had the advantage of the rule to guide the choice of $k_r$ being computationally more straightforward as well as offering better predictive performance.

\section{Application to Bipolar Disorder Data}
\label{sec:realdata}
A performance comparison similar to that described in the previous section was performed using real SNP data taken from two GWAS of Bipolar Disorder (BD). BD is a complex neurobehavioral phenotype, characterised by episodes of mania and depression. The lifetime prevalence of BD is estimated to be in the region of 1 \%. The heritability of BD has been estimated to be as high as 85 \% [\cite{McGuffin:2003p3268}]. A number of loci have been identified as associated with BD; however replication studies have not always been successful [\cite{Alsabban:2011p3269}]. It is thought that many genes of small effect contribute to the liability to develop BD. This hypothesis has been offered as an explanation for the underwhelming findings from BD GWAS [\cite{Serretti:2008p3267}]. 

We use data from the WTCCC [\cite{WTCCC2007}] to construct models to predict BD status. We then evaluate these models using an independent test data set taken from the Genetic Association Information Network [\cite{GAIN}].

Before the model was fitted and evaluated, data were pre-processed and quality control checks were performed following the documented procedures accompanying each data set. Briefly, individuals and genotype calls that had been identified as poor quality by the WTCCC were removed from the data. Missing genotypes were imputed using Impute2 [\cite{Howie:2009p2602}], with genotypes with the highest posterior probability being used in the analysis.  For the GAIN data, only individuals with European ancestry and unambiguous phenotype were used in the test data. Pre-imputation quality control involved removing one of each of pairs of individuals identified as related in the data, removing invariant SNPs, SNPs with call rate $<$ 98\%, and SNPs with Hardy-Weinberg $p$-value $<1\text{e}^{-4}$. Individuals identified as outliers by the program EIGENSTRAT  [\cite{Price:2006p53}] were removed from their respective datasets. 
 
In order to evaluate the predictive models, it was necessary to have the same predictors (SNPs) in both the training and test data sets. Approximately 300,000 autosomal SNPs that were common to both datasets were used in the analysis. PLINK v1.07 [\url{http://pngu.mgh.harvard.edu/~purcell/plink/},  \cite{Purcell:2007p2603}] was used for pre-imputation quality control and data preparation steps.

Having obtained directly typed and imputed SNPs such that we had the same SNPs in the two datasets, predictive models were fitted. In these data with a binary outcome, the logistic model was used to describe the relationship between genotypes and disease status. 

In performing variable selection followed by multiple regression, instead of including a pre-defined proportion of all predictor variables in the multiple regression model, we chose a significance threshold ($p$-value cutoff) for a variable to be included. We evaluated predictive performance at a range of $p$-value thresholds. In HyperLasso regression, as before, we fixed the shape parameter as 3.5 and chose the penalty parameter using tenfold cross-validation.

Regression coefficients were estimated using RR, with shrinkage parameter $k_r$ as in Equation \eqref{eq:kr}. Results are presented with $r$ chosen such that the degrees of freedom for variance of the resultant model is equal to the number of components used in the computation of $k_r$. In order to prevent local regions of high linkage disequilibrium (LD) overwhelming the principal components, the training data were thinned to 1 SNP every 100kb before choosing the number of principal components and computing $k_r$. This thinning of the data was evaluated in the simulation studies in the previous section, but thinning did not affect predictive performance in that case (results not shown). Fitted coefficients were subsequently estimated on the full set of SNPs.

Results comparing the predictive performance of our proposed method with that of models based on univariate tests of significance and models fitted using HyperLasso regression are presented in Table \ref{Tab:BD}. In models fitted using univariate variable selection followed by multiple regression, relaxing the significance threshold for inclusion of a SNP in the model quickly lead to more SNPs reaching the threshold than there are observations in the data. With more predictors than observations, a multiple regression model cannot be fitted. Thus when using the univariate variable selection approach, we necessarily discard information in the large number of SNPs that are moderately associated with outcome. HyperLasso regression presents the problem of the choice of the two penalty parameters which control the amount of shrinkage. Choice of the parameters by cross-validation is computationally intensive, becoming unfeasibly so for large data sets such as this one. Our method has the advantage of not requiring cross-validation and offering improved predictive performance. Again we see that our proposed estimator offers good predictive performance in comparison to other regression approaches as well as having computational advantages. 

\section{Computational Issues}\label{sec:compissues}
We have developed an R package, \verb|ridge|, which implements our method. This package addresses the computational challenges that arise when fitting RR models to high-dimensional data such as genome-wide SNP data. For data sets that are too large to read into R, code written in C is provided and the corresponding R functions take file names as arguments. This circumvents the need to read large datasets into the R workspace whilst retaining a user-friendly interface to the code.

Our method requires a number of matrix operations and linear algebra functions which can be computationally costly for large matrices. When a compatible graphical processing unit (GPU) is available, our software makes use of it, dramatically reducing computational time. NVIDIA CUDA Architecture [\cite{citeulike:10005504}] is used to access the GPU and CULA [\cite{Humphrey:2010p3192}] is used for the matrix operations and linear algebra functions. 

Logistic RR is performed using the CLG algorithm [\cite{Genkin:2007p1152}]. CLG is a cyclic coordinate descent algorithm which updates each coefficient in turn, holding the other coefficients fixed until convergence is reached. This removes the need to repeatedly manipulate an entire data matrix at once and makes logistic RR feasible even when the data contain hundreds of thousands of predictor variables. For details of the CLG algorithm, see Appendix \ref{CLG}.

\section{Discussion} In this paper we have introduced a semi-automatic method to guide the choice of shrinkage parameter in ridge regression. Our method is particularly useful when the regression problem comprises more predictor variables than observations, a situation that often arises in genetic data. This is because existing ridge estimators, such as that proposed by \cite{Hoerl1975} cannot be computed in such settings.

Disease risk prediction using genetic information remains a challenging problem due to the high dimensionality and correlation structure of the data. RR is a technique that addresses these difficulties and has been shown to offer good predictive performance. Our method facilitates the application of RR to genetic data for the construction of prediction models.

Our method has computational and practical advantages over competing methods. Because the choice of shrinkage parameter is semi-automatic, our method does not require computationally intensive cross-validation procedures such as those required by HyperLasso regression. Nor is determination of causal variables necessary, as is the case when selecting predictor variables to include in a multivariate model fitted using OLSR.

Using simulation studies, we demonstrate that our method outperforms HyperLasso regression in terms of prediction error when data comprise more predictors than observations and there are many causal variables with small effects, a situation that is representative of genetic data. We demonstrate the good predictive performance our method by using data from two genome-wide association studies to construct and evaluate a prediction model.

We have developed an R package [\cite{R2007}], \verb|ridge|, which implements our method. By taking file names as arguments, the R package can handle hundreds of thousands of predictors and thousands of observations. By default, \verb|ridge| computes the ridge parameter using our proposed method, but it flexibly allows the user to specify the ridge parameter or the number of components to use to compute it. Because RR is a regression approach, the can be extended to include additional, non-genetic covariates. For example, clinical information or PCs to correct for population structure could be included. Such covariates can be accommodated by our package, where they may be penalised or not penalised. It would be possible to extend the approach to investigate higher-order interaction terms, albeit at an increased computational cost. Given the large number of predictor variables in a GWAS, it may be necessary to perform a variable selection step before investigating interaction effects. 

\appendix
\section{Logistic ridge regression by cyclic coordinate descent}\label{CLG}
In this section, we describe logistic RR and cyclic coordinate descent, the algorithm which we use to compute logistic RR coefficients.

In the logistic regression model, let $\mathbf{X}$ be an $n \times p$ matrix of predictors with rows $\mathbf{x}_i = \left( x_{i1}, \dots, x_{ip}\right)$, as in the linear regression model (Equation \eqref{eq:linmodmat}). Let $\mathbf{Y} = \left( Y_1, \dots, Y_n \right)$ be a vector of observed binary outcomes, $Y_i~\in~\left\{ 0, 1 \right\}$. In biomedical data, this setup is common. The outcome variable represents disease status with cases as 1 and controls as 0. 

The $i^\text{th}$ response $Y_i$ is a Bernoulli variable with probability of success  $\pi\left( \mathbf{x}_i \right)$. The logistic regression model relates the probability $\pi\left( \mathbf{x}_i \right)$ that the $i^\text{th}$ observation is a case to the predictor variables as
\begin{equation}\label{eq:logitprob}
\text{Pr}\left(Y_i=1|\mathbf{x}_i\right)=\pi\left( \mathbf{x}_i \right)=\frac{e^{\mathbf{x}_i\boldsymbol{\beta}}}{1+e^{\mathbf{x}_i\boldsymbol{\beta}}}
\end{equation}
where $\boldsymbol\beta$ is a vector of parameters to be estimated. 
Logistic RR estimates are obtained by maximising the log-likelihood of the parameter vector, subject to a penalty term. The penalised log-likelihood to be maximised is:
\begin{equation}
l(\boldsymbol {\beta}) = \sum_{i=1}^n Y_i \text{log}(\pi(\mathbf{x}_i)) + \sum_{i = 1}^n(1 - Y_i)\text{log}(1 - \pi(\mathbf{x}_i)) - k\|\boldsymbol{\beta}^2\|
\end{equation}
 The CLG algorithm [\cite{Zhang:2001p1165}] is a cyclic coordinate descent algorithm for penalised logistic regression. The algorithm is described in detail by \cite{Genkin:2007p1152}. The CLG algorithm is initiated by setting all of the coefficient estimates to an initial value. Then, each coefficient is updated in turn whilst holding the rest fixed. This has the advantage of avoiding the need for the inversion of large matrices. Convergence is checked after each round of updating all of the coefficients. 
 
In the CLG algorithm, cases are code as $Y_i = 1$ and controls as $Y_i = -1$. Finding the updated coefficient, $\beta_j^\text{new}$ that maximises the log-likelihood whilst keeping the other parameters fixed is equivalent to finding the $z$ that minimizes
 \begin{equation}
  g(z) = \left( \sum_{i = 1}^n f\left( r_i + (z - \beta_j)x_{ij}y_{i}\right) \right) + \frac{z^2}{2\tau}
 \end{equation}
 where $\tau = \frac{1}{2k}$ and the $r_i = \boldsymbol{\beta}'\mathbf{x}_iy_i$ are computed using the current value of $\boldsymbol\beta$ and so are treated as constants. $f (r) = \text{ln}(1 + \text{exp}(-r))$, and penalty terms not involving $z$ are constant and therefore omitted. 

The $\beta_j^\text{new}$ that gives the minimum value of $g(\cdot)$ does not have a closed form, so each component-wise update requires an optimization process. \cite{Zhang:2001p1165} use a modification of Newton's method in computing the component-wise updates. The proposed updates are adaptively bounded to prevent large updates in regions where a quadratic is a poor approximation to the objective. Following \cite{Genkin:2007p1152} we use as the proposed update:
 \begin{equation}\label{tentUpdate}
 \Delta \nu_j = \frac{\sum_{i = 1}^n (x_{ij}y_i) / (1 + \text{exp}(r_i)) - \beta_j / \tau}  {\sum_{i = 1}^nx_{ij}^2F(r_i, \Delta_j|x_{ij}|) + 1 / \tau}
 \end{equation}
 \cite{Genkin:2007p1152} use 
 \begin{equation}
F(r, \delta) = \left\{ \begin{array}{l l}
    						0.25 & \text{if } |r| \leq \delta \\
    						\frac{1}{2 + \text{exp}(|r| - \delta) + \text{exp}(\delta - |r|)} & \text{otherwise}  \\
  					\end{array}\right. 
 \end{equation}
but other functions can be used [\cite{Zhang:2001p1165}]. We then apply the trust region restriction:
 \begin{equation}
 \Delta_j^\text{new} = \text{max} (2 |\Delta \beta_j|, \Delta_j / 2)
 \end{equation}
  
  to give the actual update:
 \begin{equation}
 \Delta \beta_j = \left\{ \begin{array}{l l}
    						-\Delta_j & \text{if } \Delta \nu_j < -\Delta_j \\
    						\Delta \nu_j & \text{if } -\Delta_j \leq \Delta \nu_j \leq \Delta_j  \\
						\Delta_j & \text{if } \Delta_j < \Delta \nu_j  \\
  					\end{array}\right. 
 \end{equation}
Convergence is declared when $(\sum_{i = 1}^n | \Delta r_i |) / (1 + \sum_{i=1}^n r_i) < \epsilon$, where $\sum_{i = 1}^n|\Delta r_i|$ is the sum of the changes in the linear scores once all the coefficients have been updated, and $\epsilon$ is a user-specified tolerance. 
 The CLG algorithm is summarized in Algorithm \ref{CLGalg}.

\begin{algorithm}                      
\caption{CLG  [\cite{Zhang:2001p1165, Genkin:2007p1152}] }          
\label{CLGalg}                           
\begin{algorithmic}                    
\REQUIRE $\beta_j \gets 0, \Delta_j \gets 1$ for $j = 1, \dots, p$; $r_i \gets 0$ for $i = 1, \dots, n$
\WHILE {} 
\FOR {$j = 1 \dots p$}
\STATE compute tentative step $\Delta \nu_j $ (Equation \eqref{tentUpdate}). 
\STATE $\Delta\beta_j \gets \text{(\text{max}())}$ (Limit step to trust region)
\STATE $\Delta r_i \gets \Delta \beta_j x_{ij}y_i$, $r_i \gets r_i + \Delta r_i$ {\bf for} $i = 1, \dots, n$
\STATE $\beta_j \gets \beta_j + \Delta \beta_j$
\STATE $\Delta_j \gets \text{max}(2|\Delta \beta_j|, \Delta_j / 2)$ (update size of trust region)
\ENDFOR
\STATE {\bf while} $(\sum_{i = 1}^n | \Delta r_i |) / (1 + \sum_{i=1}^n r_i) > \epsilon$
\ENDWHILE
\end{algorithmic}
\end{algorithm}
\clearpage \newpage
\section{Definitions of Degrees of Freedom in penalised regression models}\label{DegOfFAppendix}
In RR:
\begin{equation}
\label{eq:app}
\text{tr}(\mathbf{H}\mathbf{H}') \leq \text{tr}(\mathbf{H}) \leq \text{tr}(2\mathbf{H} - \mathbf{H}\mathbf{H}') \end{equation}
\begin{proof}
\begin{align*}
\mathbf{H} &= \mathbf{X} \left(\mathbf{X}'\mathbf{X} + k\mathbf{I}\right) \mathbf{X}' \\
&= \mathbf{U}\mathbf{D}\mathbf{V}' \left(\mathbf{V} \mathbf{D}^2 \mathbf{V}' + k\mathbf{I}\right) \mathbf{V}\mathbf{D}'\mathbf{U}' \\
&= \mathbf{U}\mathbf{D}\mathbf{V}' \left(\mathbf{V} ( \mathbf{D}^2 + k)\mathbf{V}'  \right) \mathbf{V}\mathbf{D}'\mathbf{U}' \\
&= \mathbf{U} \left[ \mathbf{D}^2 / (\mathbf{D}^2 + k) \right]  \mathbf{U}' \\
\end{align*}

$\text{tr}(\mathbf{H})$ is the sum of the $t$ diagonal elements of $\mathbf{H}$. Each element is less than or equal to 1. $\text{tr}(\mathbf{HH'})$ is also the sum of $t$ diagonal elements, this time of $\mathbf{HH'}$, and each of which is the square of the corresponding diagonal element of $\mathbf{H}$. These diagonal elements each take a value between 0 and 1, thus the sum of their squares is less than or equal to the sum of the original elements. A similar argument holds for the diagonal elements of $2\mathbf{H} - \mathbf{H}\mathbf{H}'$, where the sum is greater than or equal to the sum of the diagonal elements of $\mathbf{H}$:
\begin{align*}
\text{trace}(\mathbf{H}) &= \sum_{j=1}^t \lambda_j^2 / (\lambda_j^2 + k) \quad t = \text{min}(n, p) \\
\text{trace}(\mathbf{H}\mathbf{H}') &= \sum_{j=1}^t \lambda_j^4 / (\lambda_j^2 + k)^2 \\
\text{trace}(2\mathbf{H} - \mathbf{H}\mathbf{H}') &= \sum_{j=1}^t \lambda_j^2 (\lambda_j^2 + 2k) / (\lambda_j^2 + k)^2
\end{align*}
\end{proof}
{\sc Corollary: }For a fixed value of the degrees of freedom, $k_{\mathbf{H}\mathbf{H}'} < k_\mathbf{H} < k_{2\mathbf{H} -\mathbf{H}\mathbf{H}'}$ where $k_{\mathbf{H}\mathbf{H}'}$ is $k$ such that $\text{tr}(\mathbf{H}\mathbf{H}') = r$, $k_\mathbf{H}$ is $k$ such that $\text{tr}(\mathbf{H}) = r$ and $k_{2\mathbf{H} - \mathbf{H}\mathbf{H'}}$ is $k$ such that $\text{tr}(2\mathbf{H} - \mathbf{H}\mathbf{H}') = r$ (for the same value of $r$ in all three cases).
\begin{proof}
We seek $k_{\mathbf{H}}$ and $k_{\mathbf{H}\mathbf{H}'}$ such that:
\begin{align*}
&\sum_{j=1}^t \lambda_j^2 / (\lambda_j^2 + k_\mathbf{H})  = \sum_{j=1}^t \lambda_j^4 / (\lambda_j^2 + k_{\mathbf{H}\mathbf{H}'})^2 = r\\
\end{align*}
 For each diagonal element $j = 1 \dots t$:
 \begin{align*}
\lambda^2 ( \lambda^2 + k_{\mathbf{H}\mathbf{H}'} )^2 &= \lambda^4 (\lambda^2 + k_{\mathbf{H}} ) \\
\end{align*}
Which simplifies to 
\begin{equation*}
(2 + \frac{1}{\lambda^2})k_{\mathbf{H}\mathbf{H}'} = k_\mathbf{H}
\end{equation*}
$(2 + \frac{1}{\lambda^2}) > 0$ so $k_{\mathbf{H}} > k_{\mathbf{H}\mathbf{H}'}$

An analogous argument shows that $k_{2\mathbf{H} - \mathbf{H}\mathbf{H'}} > k_{\mathbf{H}}$.
\end{proof}
 The larger the value of $k$, the further the ridge estimates are from the OLS estimates, as expressed in Equation \eqref{eq:ak}. This relationship holds when the ridge estimates are returned to the orientation of the data by the relationship in Equation \eqref{eq:ahattobhat}.

In RR with $k > 0$, the three definitions of degrees of freedom follow the inequalities given in \eqref{eq:app}. For each of the definitions, it is possible to choose the ridge parameter such that the degrees of freedom equal some specified value. Thus, choosing $k$ such that $\text{tr}(\mathbf{H}\mathbf{H}') = r$ (among the three definitions of degrees of freedom) results in regression coefficient estimates that are closest to the OLS estimates. 
 
\clearpage \newpage
\section{Tables and figures}

\begin{table}[h]
\begin{adjustwidth}{-1in}{-1in}
\caption{Four simulation scenarios used in the evaluation of the bias-variance decomposition. The simulation scenarios are taken from \cite{Zou:2005p597}}
\label{tab:ZouHastie}
\begin{tabular} {c l l l l}
\
&&&&\\
scenario	&$n$		&$p$		&$\boldsymbol\beta$				&Structure of $\mathbf{X}$		\\ \hline
(1)	&100		&8			&$\left( 3, 1.5, 0, 0, 2, 0, 0, 0 \right)$	&	$\text{corr}\left( i, j \right) = 0.5^{|i - j|}$					\\
&&&&\\
(2)	&100		&8			&$0.85$ for all $j$					&		$\text{corr}\left( i, j \right) = 0.5^{|i - j|}$				\\
&&&&\\
(3)	&50		&40		&$\beta_j = \left\{ \begin{array}{l l}
    						0 & \quad j = (1, \dots, 10, 21, \dots, 30)\\
    						1 & \quad j = (11, \dots, 20, 31, \dots, 40) \\
  					\end{array}\right. $						&	$\text{corr}\left( i, j \right) = 0.5$ for all $i$ and $j$		\\
&&&&\\
(4)	&50		&40		&$\beta_j = \left\{\begin{array}{l l}
    0 & \quad j = (1, \dots, 15)\\
    1 & \quad j = (16, \dots, 40) \\
  \end{array} \right.$					&		$
\begin{array}{l l l }
\mathbf{x}_j = Z_1 + \epsilon_j^x,		&Z_1 \sim \mathcal{N}(0, 1)		&	j = 1, \dots, 5	\\
\mathbf{x}_j = Z_2 + \epsilon_j^x,		&Z_2 \sim \mathcal{N}(0, 1)		&	j = 6, \dots, 10	\\
\mathbf{x}_j = Z_3 + \epsilon_j^x,		&Z_3 \sim \mathcal{N}(0, 1)		&	j = 11, \dots, 15	\\
\mathbf{x}_j \sim \mathcal{N}(0, 1)		&								&	j = 16, \dots, 40 \\
\end{array}
$						\\		\hline
&&&&\\
\end{tabular}
 \end{adjustwidth}
\end{table}

\begin{table}[h]
\begin{adjustwidth}{-1in}{-1in}
\centering
\caption{Prediction squared error using $r$ with different proportions of variance explained. }
\label{Tab:PSEvsPropVar}
    \setlength{\extrarowheight}{3pt}
\begin{tabular}{ r c c c c c c c } \hline
& \multicolumn{5}{c}{\bf Proportion of variance explained (\%)} & \multicolumn{2}{r}{\bf RR parameter} \\ 
& 10 & 50 & 70 & 90 & MAX & CV & DofF \\
Continuous outcomes (mean PSE) & 1.24 & 1.23 & 1.23 & 1.27 & 3.20 & 1.24 & 1.23 \\
Binary outcomes (mean CE) & 0.46 & 0.47  & 0.47 & 0.47 & 0.47 & 0.47 & 0.46 \\
 \end{tabular}
 \end{adjustwidth}
 \end{table}
\vspace{1cm}

\begin{table}[h]
\begin{adjustwidth}{-1in}{-1in}
\centering
\caption{Prediction error - simulated data}
\label{Tab:SmallEffect}
    \setlength{\extrarowheight}{3pt}
\begin{tabular}{ r c c c c c c c } \hline
& \multicolumn{5}{c} {Univariate} & HLasso & RR \\
 {\% of SNPs ranked by univariate $p$-value} & 0.1\%  & 0.5\% & 1 \% & 3\% & 4\% &  & \\
Continuous outcomes (mean PSE) & 1.51 & 1.55 & 1.54 & 2.21 & 3.93 & 2.41 & 1.23 \\
Binary outcomes (mean CE) & 0.49 & 0.48 & 0.48 & 0.49 & 0.50 & 0.50 & 0.46 \\ \hline
 \end{tabular}
 \end{adjustwidth}
\end{table}

In Tables \ref{Tab:PSEvsPropVar} and \ref{Tab:SmallEffect}, CV refers to RR with the number of PCs used to compute $k_r$ chosen by cross-validation, and DofF refers to RR with the number of PCs used to compute $k_r$ chosen using $\text{tr}(\mathbf{H}\mathbf{H}') = r$. In Table \ref{Tab:SmallEffect}, in HyperLasso regression, the shape parameter was fixed to 3.5 and the penalty parameter was chosen using tenfold cross-validation, and in Ridge Regression the shrinkage parameter $k_r$ was used.

\begin{table}[h]
	\caption{Prediction error - Bipolar Disorder data. Logistic RR models were fitted on WTCCC-BD data. Results are reported as mean classification error over the GAIN-BD data.}
	\label{Tab:BD}
	\setlength{\extrarowheight}{3pt}
	\begin{tabular}{ r c c c c c } \hline
		& \multicolumn{3}{c}{Univariate} & HyperLasso & Ridge Regression \\
		\emph{$p$-value threshold}& $10^{-5}$ & $10^{-7}$ & $10^{-10}$ & & \\
		Mean classification error & 0.489 & 0.491 & 0.490 & 0.492 & 0.465 
	\end{tabular}
\end{table}
\clearpage \newpage
\vspace{6pc}


\clearpage \newpage

\bibliographystyle{imsart-nameyear}
\bibliography{Cule_AOAS_Bibliography}

\end{document}